\documentclass{elsart}

\begin{document}

\begin{frontmatter}



\title{Fractional generalization of the Ginzburg-Landau equation:
       An unconventional approach to critical phenomena in complex media}


\author{A. V. Milovanov\corauthref{cor1}}
\corauth[cor1]{Corresponding author.}
\ead{amilovan@mx.iki.rssi.ru}

\address{Space Research Institute, Russian Academy of Sciences, \\
         84/32 Profsoyuznaya street, 117997 Moscow, Russia}

\author{J. Juul Rasmussen}

\address{Department of Optics and Plasma Research,
Ris\o~National Laboratory, \\ OPL-128, DK-4000 Roskilde, Denmark}

\begin{abstract}
Equations built on fractional derivatives prove to be a powerful
tool in the description of complex systems when the effects of
singularity, fractal supports, and long-range dependence play a
role. In this paper, we advocate an application of the fractional
derivative formalism to a fairly general class of critical
phenomena when the organization of the system near the phase
transition point is influenced by a competing nonlocal ordering.
Fractional modifications of the free energy functional at
criticality and of the widely known Ginzburg-Landau equation
central to the classical Landau theory of second-type phase
transitions are discussed in some detail. An implication of the
fractional Ginzburg-Landau equation is a renormalization of the
transition temperature owing to the nonlocality present.
\end{abstract}

\end{frontmatter}

The formulation of fractional kinetics \cite{Nature} has led to a
considerable progress in our vision of complex systems at the
microscopic level. The use of fractional derivative operators $-$
unconventional analytical tools extending the familiar high-school
calculus to a broader class of mathematical objects $-$ has been
recognized in the modern physics \cite{Hilfer} owing to its
elegance and the proximity to the standard, well established
algorithms. The fractional kinetic equations dealing with
generalized derivatives in space and time incorporate in a
natural, unified way the key features of non-Gaussianity and
long-range dependence that often break down the restrictive
assumptions of locality and lack of correlations underlying the
conventional statistical mechanical paradigm. From a probabilistic
standpoint, fractional kinetics extends Gaussian stochastic
processes (i.e., Brownian random walks) by taking into account
long-range correlated events in the tail of the probability
density function. Such events dominate, for instance, L\'evy-type
processes \cite{Levy} and fractal time random walks (FTRW's)
\cite{FTRW}, the simplest model realizations accounting for
anomalous transport phenomena in turbulent media \cite{Oggi}.
Manifestations of fractional kinetics have been found in, e.g.,
point vortex flows \cite{CSF}, low confinement mode plasmas
\cite{Carreras}, non-Gaussianity of fluctuations measured in the
edge and scrape-off layer region of fusion devices \cite{Fusion},
etc. A comprehension of the essential role played by FTRW and
L\'evy statistics in the microscopic description of turbulence and
chaos stipulated fractional generalizations of the diffusion and
Fokker-Planck-Kolmogorov equations, discussed in a series of
publications \cite{Giona,PhysicaD,Saichev,Weitzner,PRE01}. Beside
the theory of turbulent diffusion, applications of fractional
kinetics concern the fractional Kramers problem \cite{Today},
relaxation in polymer systems and rebinding phenomena in proteins
\cite{Nonn}, scale-invariance and universality near a phase
transition point \cite{Halperin}, cosmic rays acceleration
\cite{PRE0101}, the dynamics of fracton excitations \cite{PRB02},
including modulational instability \cite{Uspehi} and self-focusing
of waves on fractals \cite{Self}, and many other realizations
\cite{Hilfer}. The current state of the art is summarized in
review articles \cite{Uspehi,Klafter,Reports}.

In this Letter, we advocate an application of the fractional
derivative formalism to the thermodynamics of second-type phase
transitions in the presence of a coexisting nonlocal ordering
which may influence the properties of the basic ``symmetric" phase
below the transition point. As an example, we mention highly
correlated electron liquid states, such as electron liquids with
fractionally charged excitations \cite{Fraction}, as well as
high-temperature superconducting fluid phases in copper-oxide
compounds and their derivatives \cite{Compound} where the nonlocal
ordering can be associated with the so-called ``stripy" order
\cite{Focus}. A growing evidence of stripes and of their role in
the superconducting transition at high temperatures is a hot topic
in condensed matter research \cite{Mix}.

In what follows, we imply a phase diagram which accommodates two
thermodynamically distinct phases, the ``symmetric"
(superconducting) phase below the transition temperature $T_c$,
and ``asymmetric" (normal) phase above the $T_c$. As usual, the
deviation between the symmetric and normal phases for $T
\rightarrow T_c$ is characterized by the order parameter, $\psi =
\psi (x)$, which is assumed to be an analytical function of a
1-dimensional coordinate variable, $x$.

Our study has two interconnected goals. First, we suggest that an
interaction between the $\psi$ order and a coexisting nonlocal
ordering can be characterized by a fractional generalization of
the free energy expansion near the transition point. Second, we
demonstrate that the order parameter $\psi$ obeys an
integro-differential equation which can be fairly considered as a
fractional extension of the widely known Ginzburg-Landau equation,
central to the classical Landau theory of second-type phase
transitions. This unconventional, fractional equation incorporates
in an analytically appealing way the effect of long-range
dependence posed by the underlying nonlocal ordering.

We start up with the conventional free energy expansion (Refs.
\cite{Original} and \cite{Landau})
\begin{equation}
F = F_n + \int _{-\infty}^{+\infty} dx \left[ \frac{\hbar
^2}{4m}\,|\nabla _x \psi| ^2 + a |\psi| ^2 + \frac{b}{2} |\psi| ^4
\right] \label{1}
\end{equation}
in vicinity of the critical point $T\rightarrow T_c$. The term
marked by $F_n$ denotes the contribution from the normal phase.
Equation (1) concerns with the real-space derivative $\nabla _x
\psi$ along the coordinate $x$. For the sake of simplicity, we
shall also use the notation $\psi _x ^{\prime}$ for $\nabla
_x\psi$. The coefficient $a = \alpha (T - T_c)$ in Eq. (1) is
proportional to the deviation between $T$ and $T_c$. Note that $a
= \alpha (T - T_c)$ changes sign at criticality: This behavior
does not depend on the nature of the symmetric and normal phases
and mirrors the generic features of topology of the phase diagram.
The system-specific information is contained in the parameters
$\alpha$ and $b$. The value of $\alpha > 0$ in accordance with the
fact that the symmetric phase occurs below $T_c$. The coefficient
$b > 0$ depends solely on the mass density of the material (but
not on the thermodynamic temperature $T$). The bulk distribution
of the order parameter extremizes the free energy functional in
Eq. (1), leading to the classical Ginzburg-Landau equation
\cite{Landau}
\begin{equation}
-\frac{\hbar ^2}{4m} \nabla ^2 _{~x} \psi + a \psi + b |\psi| ^2
\psi = 0. \label{2}
\end{equation}
The key issue about Eq. (2) is the infinitesimal $-$ local $-$
character of the $\psi (x)$ variation, manifest in the assumption
that only a differential contribution $\propto | \nabla _x\psi |
^2$ in the free energy density comes into play for $T\rightarrow
T_c$. Statistically, this means that the blobs of the symmetric
phase appear {\it at random} throughout the material as the
temperature $T$ approaches the critical range. The property of
randomness is explicit from the convolution of $\nabla _x \psi$
with a Gaussian $G _{\lambda} (x) = \exp [- x ^2 / \lambda] /
\sqrt{\pi\lambda}$, where $\lambda$ is the correlation
(coarse-graining) length, physically corresponding to the typical
size of the blobs:
\begin{equation}
\nabla _x\psi * G _{\lambda} (x) = \int _{-\infty}^{+\infty} dy\,
\psi _y ^{\prime}\, G _{\lambda} (x - y). \label{3}
\end{equation}
At length scales $x$ large compared to $\lambda$, the Gaussian $G
_{\lambda} (x)$ in Eq. (3) can be fairly approximated by the Dirac
delta function $\delta (x)$, yielding $\nabla _x\psi * G
_{\lambda} (x) \rightarrow \nabla _x\psi * \delta (x)$. The
convolution $\nabla _x\psi * \delta (x)$ is then exactly the local
derivative $\nabla _x\psi \equiv \psi _x ^{\prime}$.

We now turn to the order parameter $\Im = \Im (x)$ for the
coexisting nonlocal symmetry. The interaction between $\psi$ and
$\Im$ orders may be envisaged as a {\it nonrandom} appearance of
the ``superconducting" blobs for $T\rightarrow T_c$ over a broad
range of scales $x\gg\lambda$. In this connection, the parameter
$\Im (x)$ acquires the role of the Gaussian $G _{\lambda}
(x)\rightarrow \delta (x)$ in Eq. (3). Replacing $\delta (x)$ by
$\Im (x)$, one encounters the convolution $\nabla _x \psi * \Im
(x)$, which substitutes the local derivative $\nabla _x\psi *
\delta (x)$ in the free energy expansion (1). We may now {\it
postulate} the free energy expansion for a long-range correlated
thermodynamical system at criticality in the generalized form
\begin{equation}
F = F_n + \int _{-\infty}^{+\infty} dx \left[ A _{\Im}\,|\nabla _x
\psi * \Im| ^2 + a _{\Im}\,|\psi| ^2 + \frac{1}{2}\, b _{\Im}
|\psi| ^4 \right] \label{4}
\end{equation}
where $\Im = \Im (x)$ quantifies the underlying nonlocal symmetry.
The coefficients $A _{\Im}$, $a _{\Im}$, and $b _{\Im}$ introduced
in Eq. (4) replace $\hbar ^2 / 4m$, $a$, and $b$, respectively, in
the conventional free energy expansion in Eq. (1).

Our further interest is on the specific type of nonlocality
consistent with a self-similar $-$ fractal $-$ organization. The
focus on fractals \cite{Feder} is motivated by a general tendency
of complex systems to reveal, at or near a critical point,
scale-invariant dynamical properties \cite{Halperin}. The
phenomenon is often associated with the issue of {\it
self-organized criticality} \cite{Bak}. Self-organized critical
behavior, due to multiscale Josephson coupling of the
superconducting domains, was advocated for granular and
polycrystalline superconductors in Ref. \cite{Ginzburg}. In the
framework of our study, we assume the nonlocal symmetry has
fractal support considered as a Cantor set \cite{Feder} on the
1-dimensional Euclidean axis $x$. The fractal geometry of the
support appears in the power-law behavior of the order parameter
$\Im$: This behavior, in turn, may be identified with the scaling
of the two-point correlation function for the fractal distribution
\cite{Feder}:
\begin{equation}
\Im (x - y) = \Im _0 |x - y| ^{-\mu} \label{5}
\end{equation}
where $\Im _0$ is a normalization constant. The power exponent
$\mu$ in Eq. (5) can further be expressed in terms of the
Hausdorff fractal dimension $d_f$ of the Cantor set:
\begin{equation}
\mu = 1 - d_f. \label{6}
\end{equation}
By its definition \cite{Feder}, the Hausdorff dimension of a
Cantor set ranges from 0 to 1. In the latter case, the set
occupies the entire Euclidean axis $x$. Accordingly, the power
exponent $\mu$ varies from a maximal value equal 1 to a minimal
value equal 0. In view of Eq. (5), the convolution $\nabla _x \psi
* \Im (x)$ becomes
\begin{equation}
\nabla _x\psi * \Im (x) = \Im _0 \int _{-\infty}^{+\infty} dy\,
\psi _y ^{\prime}\, |x - y| ^{-\mu}. \label{7}
\end{equation}
The integration in Eq. (7) can be expressed in a suitable compact
form by using the notion of a {\it fractional derivative}
\cite{Vladim}. In fact, integrating by parts in Eq. (7), one gets
\begin{equation}
\nabla _x\psi * \Im (x) = \Im _0 \int _{-\infty}^{+\infty} dy\,
\psi (y)\, \nabla _{-y} |x - y| ^{-\mu}. \label{8}
\end{equation}
The operation $\nabla _{-y}$ applied to $|x - y|$ is equivalent
with the derivative $\nabla _{x}$ over the parameter $x$: This
derivative can then be taken out of the integral sign, yielding
\begin{equation}
\nabla _x\psi * \Im (x) = \Im _0 \nabla _{x} \int
_{-\infty}^{+\infty} dy\, \psi (y)\, |x - y| ^{-\mu}. \label{9}
\end{equation}
Splitting the integration from $-\infty$ to $+\infty$ into two
integrals, from $-\infty$ to $x$ and from $x$ to $+\infty$, and
taking into account the reflection symmetry $\psi (-y) = \psi (y)$
for the $\psi$ order, from Eq. (9) one obtains
\begin{equation}
\nabla _x\psi * \Im (x) = 2\Im _0 \nabla _{x} \int _{-\infty}^{x}
dy\, \psi (y)\, (x - y) ^{-\mu}. \label{10}
\end{equation}
Setting the normalization $2\Im _0 = 1 / \Gamma (1 - \mu)$ in Eq.
(5), we find $\nabla _x\psi * \Im (x) \equiv \nabla _{~x}
^{\mu}\psi$, where
\begin{equation}
\nabla _{~x} ^{\mu}\psi \equiv \frac{1}{\Gamma (1 - \mu)} \nabla
_{x} \int _{-\infty}^{x} dy\, \psi (y)\, (x - y) ^{-\mu}
\label{11}
\end{equation}
is exactly the Riesz definition \cite{Vladim} of the fractional
derivative of order $0 < \mu\leq 1$, and $\Gamma$ denotes the
Euler gamma function. Note that $\nabla _{~x} ^{\mu}$ is {\it
integro-differential} operator for all $0< \mu < 1$. In the
``integer" limit of $\mu \rightarrow 1$, the operation in Eq. (11)
is equivalent with the conventional first-order derivative $\nabla
_x$: The proof rests on the Abel identities discussed in Ref.
\cite{Vladimir}. In terms of fractional derivatives, the free
energy expansion in Eq. (4) reads
\begin{equation}
F = F_n + \int _{-\infty}^{+\infty} dx \left[ A _{\mu}\,|\nabla
^{\mu} _{~x} \psi| ^2 + a _{\mu}\,|\psi| ^2 + \frac{1}{2}\, b
_{\mu} |\psi| ^4 \right]. \label{12}
\end{equation}
Here we changed the subscript $\Im$ to $\mu$ everywhere in $A
_{\Im}$, $a _{\Im}$, and $b _{\Im}$. Expression (12) leads to the
issue of a {\it fractional Ginzburg-Landau equation}, as we now
proceed to show. In fact, varying the integral in Eq. (12) over
the complex conjugate $\psi ^*$ and considering $\psi$ and $\psi
^*$ as independent order parameters, we have
\begin{equation}
\delta F = \int _{-\infty}^{+\infty} dx \left[ A _{\mu}\,\nabla
^{\mu} _{~x} \psi \nabla ^{\mu} _{~x} \delta \psi ^* + a _{\mu}\,
\psi \delta \psi ^* + b _{\mu} |\psi| ^2 \psi \delta \psi ^*
\right]. \label{13}
\end{equation}
Making use of the integration-by-part formula \cite{Vladim}
\begin{equation}
\int _{-\infty}^{+\infty} dy\, \varphi _1 (y) \nabla ^{\mu}
_{~y}\varphi _2 (y) = \int _{-\infty}^{+\infty} dy\, \varphi _2
(y) \nabla ^{\mu} _{~-y}\varphi _1 (y) \label{14}
\end{equation}
from Eq. (13) one arrives at
\begin{equation}
\delta F = \int _{-\infty}^{+\infty} dx \left[ A _{\mu}\,\nabla
^{\mu} _{~-x} \nabla ^{\mu} _{~x} \psi  + a _{\mu}\, \psi + b
_{\mu} |\psi| ^2 \psi \right] \delta \psi ^* \label{15}
\end{equation}
yielding, in view of the extremum $\delta F = 0$,
\begin{equation}
A _{\mu}\,\nabla ^{\mu} _{~-x} \nabla ^{\mu} _{~x} \psi  + a
_{\mu}\, \psi + b _{\mu} |\psi| ^2 \psi = 0. \label{16}
\end{equation}
Varying the integral in Eq. (12) over $\psi$ leads to the
conjugate equation
\begin{equation}
A _{\mu}\,\nabla ^{\mu} _{~-x} \nabla ^{\mu} _{~x} \psi ^*  + a
_{\mu}\, \psi ^* + b _{\mu} |\psi| ^2 \psi ^* = 0 \label{17}
\end{equation}
which is physically identical to Eq. (16). Equation (16) can be
considered as a fractional generalization of the Ginzburg-Landau
Eq. (2). The fractional Ginzburg-Landau Eq. (16) determines the
bulk distribution of the order parameter $\psi$ in the presence of
a coexisting nonlocal symmetry, whose support is a Cantor set of
the Hausdorff dimension $d_f = 1 - \mu$. The coefficient $A
_{\mu}$ consistent with the fractional diffeo-integration in Eqs.
(12) and (16) could be defined by
\begin{equation}
A _{\mu} = \lambda ^{2\mu - 2}\frac{\hbar ^2}{4m} \label{18}
\end{equation}
where $\lambda$ is the microscopic correlation length. In the
conventional case of $\mu \rightarrow 1$, the value of $A _{\mu}$
reduces to $\hbar ^2 / 4m$.

Let us now discuss a likely form for the coefficient $a _{\mu}$ in
Eq. (16). Without loss of generality, we may admit that $a _{\mu}$
is a linear function on thermodynamic temperature $T$. (We assume
that all explicit nonlinearities have already been allocated to a
single term traced by $b_{\mu}$.) Let $a _{\mu}$ cross zero at
some $T = T_{\mu}$, i.e.,
\begin{equation}
a _{\mu} = \alpha (T - T_{\mu}). \label{19}
\end{equation}
Equation (19) is a simplest extension of the conventional formula
$a = \alpha (T - T_{c})$ central to the classical Landau theory
(Refs. \cite{Original} and \cite{Landau}). An important issue
about Eq. (19) is that the cross-over temperature $T_{\mu}$ where
$a_{\mu}$ changes sign may not coincide with the transition
temperature $T_c$ when the effects of nonlocality play a role
(i.e., when the index $\mu$ is smaller than 1). From the
standpoint of a formal derivation, the value of $T_c$ enabling the
occurrence of a nontrivial $\psi$ order in the presence of the
competing nonlocal symmetry decouples from $T_{\mu}$ owing to the
fractional nature of the Riesz operators $\nabla ^{\mu} _{~ - x}$
and $\nabla ^{\mu} _{~x}$ which integrate the response from the
heavy-tailed correlation function $\Im (x - y)$ in Eqs. (11) and
(16) and thereby contribute (along with the parameter $a_{\mu}$)
into an amplitude in front of $\psi$. (This amplitude then
balances the nonlinear term $b_{\mu} |\psi| ^2$ below the
transition point.) Physically, the deviation between $T_c$ and
$T_{\mu}$ has the sense of a characteristic energy of correlations
contained in the fractional Laplacian, $A _{\mu}\, \nabla ^{\mu}
_{~-x} \nabla ^{\mu} _{~x}$. By order of magnitude,
\begin{equation}
T_c - T_{\mu} \sim A_{\mu} / \alpha \lambda ^{2\mu} \Gamma ^2 (1 -
\mu) \label{20}
\end{equation}
where $\Gamma$ is the Euler gamma function. As $\mu$ tends to 1,
the deviation in Eq. (20) vanishes: $\Gamma (1 - \mu) \rightarrow
\infty$ leading to $T_{\mu} \rightarrow T_c$. The conventional
expression $a = \alpha (T - T_{c})$ is then recovered from Eq.
(19). In this limit, the nonlocal symmetry confines on a set whose
Hausdorff measure is zero, i.e., $d_f \rightarrow 0$ for $\mu
\rightarrow 1$. This behavior reproduces the assumptions of the
classical theory \cite{Landau}.

In the opposite limit of $\mu \rightarrow 0$, the fractal support
underlying the nonlocal symmetry extends to the whole of the
1-dimensional coordinate space whose Hausdorff measure is now
maximized to be $d_f\rightarrow 1$. This almost regular
distribution bears features enabling to associate it with a
``stripe" \cite{Focus}, a long-range ordering that seems to
mediate superconductivity in complex materials such as
copper-oxide compounds and their derivatives \cite{Mix}.

As $\mu \rightarrow 0$, the gamma function in Eq. (20) drops to a
minimal value equal $1$, enabling a considerable deviation between
$T_c$ and $T_{\mu}$. It can be conjectured that, in general, this
deviation accounts for the ``anomalous" transition temperatures
observed in high-temperature superconductors. A self-consistent
estimate of the parameter $T_c$ may then be obtained from Eq.
(20). A key step is the expression (Refs. \cite{Landau} and
\cite{Gorkov})
\begin{equation}
\alpha \sim \Upsilon \times (T_c / \varepsilon _F) \label{21}
\end{equation}
which relates the quantity $\alpha$ to the system-specific
characteristics such as electron energy at the Fermi edge,
$\varepsilon _F$ (typically, of the order of few eV). The
coefficient $\Upsilon$ in Eq. (21) depends on the nature of the
microscopic pairing mechanism (e.g., phonon, fracton, or exciton)
(see Refs. \cite{PRB02} and \cite{Blumen}). In the case of phonons
$-$ acoustic or (quasi)acoustic modes serving as an interface for
the conventional Bardeen-Cooper-Schrieffer (BCS) picture of
superconductivity in regular crystals \cite{Bardeen} $-$ the value
of $\Upsilon \sim 7$ (Refs. \cite{Landau} and \cite{Gorkov}), but
it can be much smaller (of the order of 1 or even less) in the
Little's exciton scenario, in which phonons are replaced by other
electrons \cite{Little}. Combining Eqs. (20) and (21), we find, in
the extreme of $\mu\rightarrow 0$,
\begin{equation}
T_c \sim \frac{\hbar}{2\lambda}\sqrt{\frac{\varepsilon
_F}{m\Upsilon}} \label{22}
\end{equation}
provided that $T_c \gg T_{\mu}$. Assuming, further, the
correlation length $\lambda$ is of nanoscales $-$ the range
typically found for the high-temperature superconductors
\cite{Mladen} $-$ from Eq. (22) one concludes a characteristic
transition temperature $T_c \sim 100$ K where the estimate
$\Upsilon \sim 7$ has been used. Non-BCS pairing processes
corresponding to smaller values of $\Upsilon$ would imply a
considerably higher critical temperature $T_c$, up to
room-temperature values. A feasible role of such processes in
two-dimensional electron systems with multi-scale long-range
correlations is addressed in Refs. \cite{Blumen} and \cite{Cor}. A
fractional kinetics constituent in the occurrence of the ensuing
superconducting fluid state will be the subject of a forthcoming
publication. We believe this constituent is crucial for the
unconventional superconductivity in materials with complex
microscopic organization.


A. V. M. is grateful to G. M. Zaslavsky for illuminating
discussions on different aspects of fractional kinetics. This
study was sponsored by the Science Support Foundation, INTAS Grant
03-51-3738, RFBR project 03-02-16967, and ``Scientific School"
Grant 1739.2003.2.


\end{document}